\documentclass[conference, 10pt, final]{IEEEtran}

\usepackage{enumitem}
\usepackage[nolist]{acronym} 
\usepackage{tikz}
\usepackage{bm}
\usepackage{pgfplots}
\usepgfplotslibrary{groupplots}
\usepackage{graphicx}
\usepackage{subcaption}
\usepackage{float}
\pgfplotsset{compat=newest}
\usepackage{units}
\usepackage{amsmath, amsbsy, amssymb, amsthm}
\usepackage{tikzscale}
\usetikzlibrary{arrows}
\usepackage{color}
\usepackage{epigraph}
\usepackage{circuitikz}
\usepackage{multirow}
\usepackage{booktabs}
\usepackage[plainruled]{algorithm2e}
\usepackage[group-separator={,}]{siunitx}
\definecolor{darkgreen}{rgb}{0.125,0.5,0.169}
\usetikzlibrary{shapes,arrows}
\usetikzlibrary{positioning}
\usetikzlibrary{patterns}
\usetikzlibrary{decorations.pathreplacing}
\usepackage{marvosym}
\usepackage{bbm}
\usepackage{mathtools}
\usepackage{xfrac}
\usepackage{numprint}

\tikzset{>=latex}

\begin{acronym}
 \acro{CSI}{channel state information}
 \acro{UE}{user equipment}
 \acro{UL}{uplink}
 \acro{BS}{basestation}
 \acro{TDD}{time division duplex}
 \acro{FDD}{frequency division duplex}
 \acro{ECC}{error-correcting code}
 \acro{MLD}{maximum likelihood decoding}
 \acro{HDD}{hard decision decoding}
 \acro{IF}{intermediate frequency}
 \acro{RF}{radio frequency}
 \acro{SDD}{soft decision decoding}
 \acro{NND}{neural network decoding}
 \acro{CNN}{convolutional neural network}
 \acro{ML}{maximum likelihood}
 \acro{GPU}{graphical processing unit}
 \acro{BP}{belief propagation}
 \acro{LTE}{Long Term Evolution}
 \acro{BER}{bit error rate}
 \acro{SNR}{signal-to-noise-ratio}
 \acro{ReLU}{rectified linear unit}
 \acro{BPSK}{binary phase shift keying}
 \acro{QPSK}{quadrature phase shift keying}
 \acro{AWGN}{additive white Gaussian noise}
 \acro{MSE}{mean squared error}
 \acro{LLR}{log-likelihood ratio}
 \acro{MAP}{maximum a posteriori}
 \acro{NVE}{normalized validation error}
 \acro{BCE}{binary cross-entropy}
 \acro{CE}{cross-entropy}
 \acro{BLER}{block error rate}
 \acro{SQR}{signal-to-quantisation-noise-ratio}
 \acro{MIMO}{multiple-input multiple-output}
 \acro{OFDM}{orthogonal frequency division multiplex}
 \acro{RF}{radio frequency}
 \acro{LOS}{line of sight}
 \acro{NLoS}{non-line of sight}
 \acro{NMSE}{normalized mean squared error}
 \acro{CFO}{carrier frequency offset}
 \acro{SFO}{sampling frequency offset}
 \acro{IPS}{indoor positioning system}
 \acro{TRIPS}{time-reversal IPS}
 \acro{RSSI}{received signal strength indicator}
 \acro{MIMO}{multiple-input multiple-output}
 \acro{ENoB}{effective number of bits}
 \acro{AGC}{automatic gain control}
 \acro{ADC}{analog to digital converter}
 \acro{ADCs}{analog to digital converters}
 \acro{FB}{front bandpass}
 \acro{FPGA}{field programmable gate array}
 \acro{JSDM}{Joint Spatial Division and Multiplexing}
 \acro{NN}{neural network}
 \acro{IF}{intermediate frequency}
 \acro{LoS}{line-of-sight}
 \acro{NLoS}{non-line-of-sight}
 \acro{DSP}{digital signal processing}
 \acro{AFE}{analog front end}
 \acro{SQNR}{signal-to-quantisation-noise-ratio}
 \acro{SINR}{signal-to-interference-noise-ratio}
 \acro{ENoB}{effective number of bits}
 \acro{PCB}{printed circuit board}
 \acro{EVM}{error vector mangnitude}
 \acro{CDF}{cumulative distribution function}
 \acro{MRC}{maximum ratio combining}
 \acro{MRP}{maximum ratio precoding}
 \acro{MRT}{maximum ratio transmission}
 \acro{DeepL}{deep-learning}
 \acro{DL}{downlink}
 \acro{SISO}{single-input single-output}
 \acro{SGD}{stochastic gradient descent}
 \acro{CP}{cyclic prefix}
 \acro{MISO}{Multiple Input Single Output}
 \acro{LMMSE}{linear minimum mean square error}
 \acro{ZF}{zero forcing}
 \acro{USRP}{universal software radio peripheral}
 \acro{RNN}{recurrent neural network}
 \acro{GRU}{gated recurrent unit}
 \acro{LSTM}{long short-term memory}
 \acro{NTM}{neural turing machine}
 \acro{DNC}{differentiable neural computer}
 \acro{TCN}{temporal convolutional network}
 \acro{FCL}{fully connected layer}
 \acro{MANN}{memory augmented neural network}
 \acro{RNN}{recurrent neural network}
 \acro{DNN}{dense neural network}
 \acro{FIR}{finite impulse response}
 \acro{BPTT}{back-propagation through time}
 \acro{GAN}{generative adversarial network}
 \acro{ELU}{exponential linear unit}
 \acro{tanh}{hyperbolic tangent}
 \acro{BICM}{bit-interleaved coded modulation}
 \acro{OTA}{over-the-air}
 \acro{IM}{intensity modulation}
 \acro{DD}{direct detection}
 \acro{RL}{reinforcement learning}
 \acro{SDR}{software-defined radio}
 \acro{WGAN}{Wasserstein generative adversarial network}
 \acro{BMD}{bit-metric decoding}
 \acro{BMI}{bit-wise mutual information}
 \acro{LDPC}{low-density parity-check}
 \acro{IDD}{iterative demapping and decoding}
 \acro{JSD}{Jensen-Shannon divergence}
 \acro{MMSE}{minimum mean square error}
 \acro{FFT}{fast Fourier transform}
 \acro{IFFT}{inverse fast Fourier transform}
 \acro{QAM}{quadrature amplitude modulation}
 \acro{EMD}{earth mover's distance}
 \acro{TDL}{tapped delay line}
 \acro{KL}{Kullback-Leibler}
 \acro{PRACH}{physical random access channel}
 \acro{URLLC}{ultra-reliable low-latency communication}
 \acro{ANOMA}{asynchronous non-orthogonal multiple access}
 \acro{FEC}{forward error correction}
 \acro{PAPR}{peak-to-average power ratio}
 \acro{APP}{a posteriori probability}
 \acro{COTS}{commercial off-the-shelf}
 \acro{PLL}{phase locked loop}
 \acro{STO}{sampling time offset}
 \acro{SFO}{sampling frequency offset}
 \acro{CFO}{carrier frequency offset}
 \acro{CPO}{carrier phase offset}
 \acro{CSI}{channel state information}
 \acro{GNSS}{global navigation satellite system}
 \acro{ELAA}{extremely large aperture array}
 \acro{UE}{user equipment}
 \acro{DICHASUS}{\underline{Di}stributed \underline{Cha}nnel \underline{S}ounder by \underline{U}niversity of \underline{S}tuttgart}
 \acro{JCAS}{Joint Communication and Sensing}
 \acro{AoA}{angle of arrival}
 \acro{REFTX}{reference transmitter}
 \acro{MOBTX}{mobile transmitter}
 \acro{AARX}{antenna array receiver}
 \acro{LO}{local oscillator}
\end{acronym}

\definecolor{mittelblau}{RGB}{0, 126, 198}
\definecolor{violettblau}{cmyk}{0.9, 0.6, 0, 0}
\definecolor{rot}{RGB}{238, 28 35}
\definecolor{apfelgruen}{RGB}{140, 198, 62}
\definecolor{gelb}{RGB}{1, 221, 0}
\definecolor{orange}{RGB}{244, 111, 33}
\definecolor{pink}{RGB}{237, 0, 140}
\definecolor{lila}{RGB}{128, 10, 145}
\definecolor{hellgrau}{RGB}{224, 224, 224}
\definecolor{mittelgrau}{RGB}{128, 128, 128}
\definecolor{dunkelgrau}{RGB}{80,80,80}
\definecolor{anthrazit}{RGB}{19, 31, 31}

\definecolor{bgorange}{HTML}{fcc0a7}
\definecolor{bggreen}{HTML}{ccebb9}

\DeclareMathOperator*{\argmax}{arg\,max}
\DeclareMathOperator*{\argmin}{arg\,min}
\DeclareMathOperator*{\diag}{diag}

\IEEEoverridecommandlockouts

\begin{document}

\title{Geometry-Based Phase and Time Synchronization for Multi-Antenna Channel Measurements}

\author{\IEEEauthorblockN{Florian Euchner, Phillip Stephan, Marc Gauger, Stephan ten Brink \\}

\IEEEauthorblockA{
Institute of Telecommunications, Pfaffenwaldring 47, University of  Stuttgart, 70569 Stuttgart, Germany \\ \{euchner,stephan,gauger,tenbrink\}@inue.uni-stuttgart.de
}

}

\maketitle

\begin{abstract}
    Synchronization of transceiver chains is a major challenge in the practical realization of massive MIMO and especially distributed massive MIMO.
    While frequency synchronization is comparatively easy to achieve, estimating the carrier phase and sampling time offsets of individual transceivers is challenging.
    However, under the assumption of phase and time offsets that are constant over some duration and knowing the positions of several transmit and receive antennas, it is possible to estimate and compensate for these offsets even in scattering environments with multipath propagation components.
    The resulting phase and time calibration is a prerequisite for applying classical antenna array processing methods to massive MIMO arrays and for transferring machine learning models either between simulation and deployment or from one radio environment to another.
    Algorithms for phase and time offset estimation are presented and several investigations on large datasets generated by an over-the-air-synchronized channel sounder are carried out.
\end{abstract}

\acresetall

\section{Introduction}
Massive \ac{MIMO} is widely accepted to be a crucial technology for increasing the spectral efficiency of future cellular wireless systems through spatial multiplexing.
Distributed massive \ac{MIMO}, where antennas are distributed along building facades or even across multiple buildings, is of particular interest thanks to the thereby greatly improved spatial macro-diversity \cite{ngo2015cell}.
For these deployments, frequency, time and phase synchronization of the distributed transceiver chains is an important subject.
Frequency synchronization is usually straightforward to achieve using \ac{OTA} reference signals, through the backhaul network or based on other accurate frequency references.
Phase and sampling times, however, are usually only approximately synchronous over certain timeframes and only down to constant, antenna-specific offsets, which makes calibration challenging.

These offsets, if they occur reciprocally for both uplink and downlink channel, are not an issue for massive MIMO operated in \ac{TDD} mode.
Since they are constant, they are also irrelevant for most deep learning-based applications of channel data (e.g., localization), as neural networks can easily learn to compensate for them.
However, phases calibrated in absolute terms are essential for classical array processing techniques (e.g., \ac{AoA} estimation) and for transferring deep learning models between different environments or between simulation and reality.

\begin{table}
    \centering
    \begin{tabular}{r | l}
        $\mathbf A$, $\mathbf s$ & Bold letters: Uppercase for matrices, lowercase for vectors \\
        $\ell, N$ & Italic uppercase or lowercase letters: Scalars \\
        $\mathbf A_{\ell:}$, $\mathbf A_{:d}$ & The $\ell$\textsuperscript{th} row /  $d$\textsuperscript{th} column of $\mathbf A$ as a column vector \\
        $\lVert \mathbf A \rVert_\mathrm{F}$ & The Frobenius norm of matrix $\mathbf A$ \\
        $\lVert \mathbf v \rVert$ & The Euclidean norm of vector $\mathbf d$ \\
        $\mathbf A^\mathrm{H}$ & The conjugate transpose of matrix $\mathbf A$ \\
        $\mathbf a \odot \mathbf b$ & Hadamard (elementwise) product of vectors $\mathbf a$ and $\mathbf b$ \\
        $\mathbf A_{:d}^\mathrm{-1}$ & Elementwise inverse of vector $\mathbf A_{:d}$
    \end{tabular}
    \caption{Symbols and notations used in this paper}
    \vspace{-0.2cm}
    \label{tab:notations}
\end{table}

Phase and time offset calibration in antenna arrays is a well-known field of study with a wide range of applications, from radio and audio signal processing to seismology.
For example, \cite{pierre1991experimental} proposes a calibration method that can estimate ultrasonic sensor array phase and gain offsets as well as mutual coupling errors based on a set of known transmitter locations in a well-controlled environment, but assumes concurrent transmission from multiple locations.
In addition to estimating gain, phase and mutual coupling errors, \cite{ng1996sensor} also accounts for sensor location errors within a planar area, but assumes known transmit signal phases.
In \cite{friedlanderweiss}, gain, phase and mutual coupling uncertainties are corrected in a self-calibration step without the need for known transmitter locations, but only within a single antenna array.
The contribution of this paper is to solve the sensor array calibration problem under conditions that are typical for distributed channel sounder measurements:
\begin{itemize}
    \item There are multiple antenna arrays distributed in space and their location is precisely known.
    \item Synchronization has been achieved except for a constant, antenna-specific phase and sampling time offset.
    \item Channel measurements with unknown starting phase for multiple known transmitter locations are available.
\end{itemize}
We propose to estimate \ac{LO} phase and sampling time offsets in the context of an \ac{OFDM} system with the help of the known antenna locations based on the assumption of strong \ac{LoS} propagation paths.
We show that geometry-based calibration is possible despite the distributed system architecture and the unknown transmitter phases.
In Sec. \ref{sec:dichasus}, we introduce our channel sounder.
The calibration problem is stated in Sec. \ref{sec:problem} and an estimation procedure based on one of two different algorithms is derived in Sec. \ref{sec:offsetestimation}.
Sec. \ref{sec:experiments} presents some experimental results and finally, the performance of the two estimation algorithms at low \acp{SNR} is compared in Sec. \ref{sec:performancecomparison}.
Throughout the paper, we use the notations defined in Tab. \ref{tab:notations}.




\section{Our Channel Sounder}
\label{sec:dichasus}
\begin{figure}
    \centering
    \includegraphics[width=0.9\columnwidth]{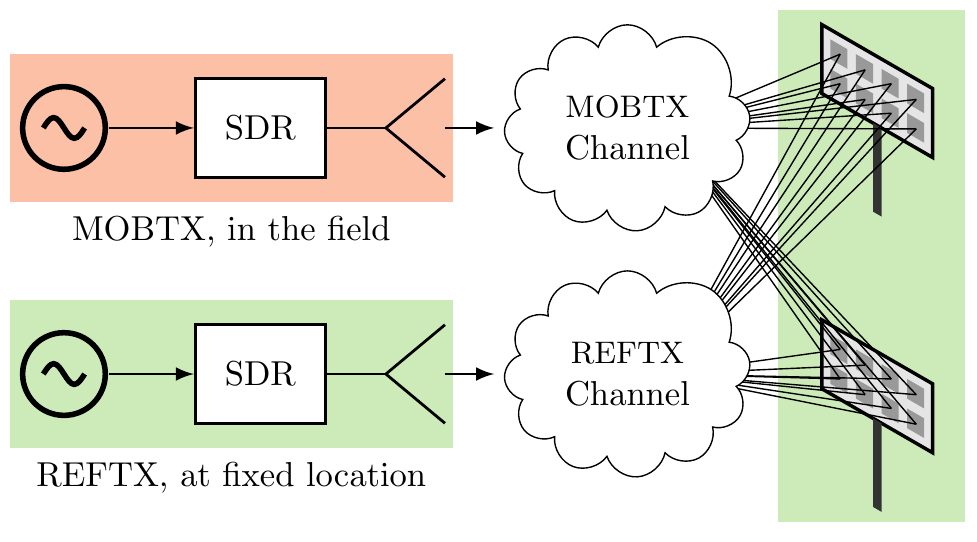}
    \caption{Overview over the DICHASUS system. Background colors indicate frequency-synchronous domains.}
    \label{fig:system}
\end{figure}

To tackle some of the challenges of distributed massive \ac{MIMO}, we developed a distributed massive \ac{MIMO} channel sounder called \ac{DICHASUS}, presented in \cite{dichasus2021}.
\ac{DICHASUS} can capture large \ac{CSI} datasets\footnote{All datasets are available at\newline\texttt{https://dichasus.inue.uni-stuttgart.de}}, containing frequency-domain channel transfer functions.
The transfer functions are measured between multiple, potentially spatially distributed, receiver antennas and a \ac{MOBTX} in the field.
Channel measurements are tagged with accurate ``ground truth'' data including antenna locations.
Positions are obtained using a robotic total station (tachymeter) with an accuracy on the order of a few millimeters or less.
\ac{DICHASUS} uses \ac{OTA} frequency, time and phase calibration for the receivers, based on a \ac{REFTX} broadcast.
\ac{REFTX} and \ac{MOBTX} operate continuously and are multiplexed in frequency.
Thanks to real-time adjustments and a software-based postprocessing step, the \ac{AARX} chains are almost perfectly synchronous to their respective received \ac{REFTX} signal in both \ac{LO} and sampling clock frequency and phase.
However, as illustrated in Fig. \ref{fig:system}, this does not necessarily imply that they are synchronous in absolute terms, since the \ac{OTA} synchronization broadcast passes through different, unknown channels between \ac{REFTX} and each \ac{AARX} antenna, leading to the aforementioned antenna-specific constant phase and time offsets.

The datasets measured by our \ac{DICHASUS} channel sounder contain channel coefficient estimates for each of the $N_\mathrm{sub}$ \ac{OFDM} subcarriers of the \ac{MOBTX} signal measured at all $L$ \ac{AARX} antennas and at $D$ different time instances.
Furthermore, the 3-dimensional \ac{MOBTX} position $\mathbf x_d \in \mathbb R^3$ at each time instance $d$ is recorded.
The location (and orientation) of each receive antenna $\mathbf y_\ell \in \mathbb R^3, ~ \ell \in \{ 1, \ldots, L \}$ can be reconstructed from metadata.
For the following derivations, it is beneficial to split the dataset by subcarrier into matrices $\mathbf R[n] \in \mathbb C^{L \times D}$ containing the channel coefficients for all antennas and time instances for one particular subcarrier $n$.


\section{Problem Statement}
\label{sec:problem}
\subsection{Timing Impairment Model}
As long as the sampling time offsets $t_{\mathrm{off}, \ell}$ for antennas $\ell \in \{ 1, \ldots, L \}$ are sufficiently small, their effect in an \ac{OFDM} system can be modelled simply as a linear phase rotation over the subcarrier index $n$.
With $g_{\ell}[n] \in \mathbb C$ denoting the phase and gain offset experienced by antenna $\ell$ for subcarrier $n$, $\varphi_{\mathrm{off}, \ell} \in [0, 2\pi)$ as the phase offset of the lowermost subcarrier at $n = 0$, $t_{\mathrm{off}, \ell}$ as the sampling time offset of antenna $\ell$ and $\Delta f_\mathrm{sub}$ being the \ac{MOBTX} subcarrier spacing, this linear phase shift can be expressed as
\begin{equation}
    \arg(g_{\ell}[n]) = \varphi_{\mathrm{off}, \ell} + 2 \pi n t_{\mathrm{off}, \ell} \Delta f_\mathrm{sub} + z_\ell.
    \label{eq:sto}
\end{equation}
In (\ref{eq:sto}), $z_\ell$ are residual error terms.
The objective of our calibration is to obtain estimates for $\varphi_{\mathrm{off}, \ell}$ and $t_{\mathrm{off}, \ell}$ from $\mathbf R[n], ~ n \in \{ 1, \ldots, N_\mathrm{sub} \}$ and $\mathbf x_d, ~ d \in \{ 1, \ldots, D \}$.



\subsection{Antenna Model}
\begin{figure}
    \centering
    \begin{subfigure}[t]{0.4\columnwidth}
        \centering
        \includegraphics[height=1.12\textwidth]{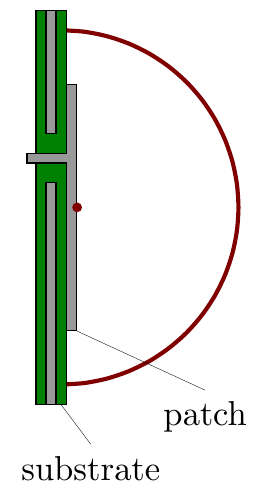}
        \caption{Ideal, isotropic case}
    \end{subfigure}
    \begin{subfigure}[t]{0.4\columnwidth}
        \centering
        \includegraphics[height=1.12\textwidth]{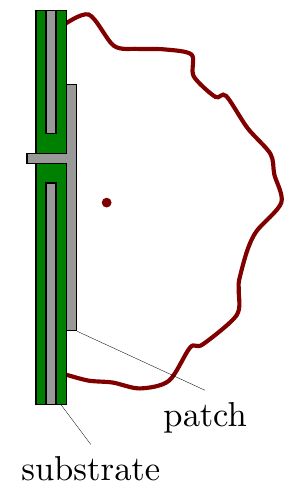}
        \caption{Realistic, phase errors}
    \end{subfigure}
    \caption{Comparison of ideal and realistic patch antenna: Phase center location (red dot) and equiphase surface (red line)}
    \label{fig:antennamodel}
\end{figure}

Our calibration method is based on comparing measured channels to ideal line-of-sight channels computed based on transmit and receive antenna locations $\mathbf x_d$ and $\mathbf y_\ell$.
It is therefore dependent on the antenna type and the antenna model.
The measurements analyzed in this work were captured using probe-fed microstrip patch antennas, shown in Fig. \ref{fig:antennamodel}, as receive antennas and dipole antennas at the transmitters.
For simplicity, we model all antennas as isotropic radiators, with the phase center at the center of the patch or dipole.
This idealized model introduces several errors, most notably:
\begin{itemize}
    \item The radiation pattern of the antenna is neglected, leading to unrealistic expected amplitudes.
    \item The true phase center of the antenna might actually be at a different location (e.g., above the patch surface).
    \item The equiphase surface of the antenna, i.e., the set of all points at which the phase of a radiated wave is equal at a given time instance, may not be a perfect sphere.
\end{itemize}
The latter two errors, illustrated in Fig. \ref{fig:antennamodel}, are especially problematic for our calibration approach, which is based on computing expected phases at the phase center.
More detailed antenna models, if available, can therefore enhance our proposed calibration.

\section{Offset Estimation Procedure}
\label{sec:offsetestimation}
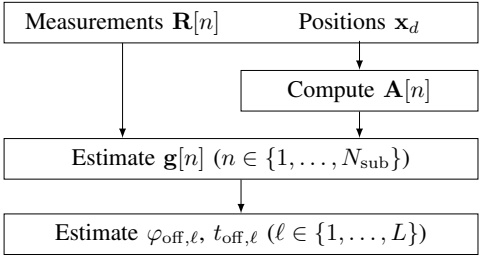
\begin{figure}
    \centering
    \scalebox{0.9}{
        \begin{tikzpicture}
            \node (dataset) [draw, minimum height = 0.6cm, minimum width = 7cm] at (0, 0) {};
            \node (R) [left = 1.75cm of dataset.center, anchor = center] {Measurements $\mathbf R[n]$};
            \node (x) [right = 1.75cm of dataset.center, anchor = center] {Positions $\mathbf x_d$};
            \node (computea) [draw, minimum height = 0.6cm, minimum width = 3.5cm, below = 0.43cm of x] {Compute $\mathbf A[n]$};
            \node (computeg) [draw, minimum height = 0.6cm, minimum width = 7cm, below = 1.4cm of dataset] {Estimate $\mathbf g[n]$ ($n \in \{ 1, \ldots, N_\mathrm{sub} \}$)};
            \node (computephasetime) [draw, minimum height = 0.6cm, minimum width = 7cm, below = 0.5cm of computeg] {Estimate $\mathbf \varphi_{\mathrm{off}, \ell}$, $t_{\mathrm{off}, \ell}$ ($\ell \in \{ 1, \ldots, L \}$)};
    
            \draw [-latex] (R.south) -- (R.south|-computeg.north);
            \draw [-latex] (x) -- (computea);
            \draw [-latex] (computea.south) -- (computea.south|-computeg.north);
            \draw [-latex] (computeg.south) -- (computephasetime.north);
        \end{tikzpicture}
    }
    \caption{Overview over all processing steps}
    \label{fig:processingoverview}
\end{figure}

Estimating the aforementioned phase and time offsets $\varphi_{\mathrm{off}, \ell}$ and $t_{\mathrm{off}, \ell}$ is based on a three-step procedure shown in \mbox{Fig. \ref{fig:processingoverview}}:
First, the \emph{ideal} expected channel coefficient matrices $\mathbf A[n]$ are computed based on $\mathbf x_d$ in Sec. \ref{sec:computingideal}. Then the subcarrier-specific phase and gain error vectors $\mathbf g[n]$ are estimated in Sec. \ref{sec:subcarrierspecific}.
Finally, the overall phase and time offsets $\varphi_{\mathrm{off}, \ell}$ and $t_{\mathrm{off}, \ell}$ are estimated based on $\mathbf g[n]$ in Sec. \ref{sec:estimatephasetime}.

\subsection{Computing Ideal Channel Coefficient Matrices $\mathbf A[n]$}
\label{sec:computingideal}
Neglecting constant factors, the expected channel coefficient for antenna $\ell$, time instance $d$ and subcarrier index $n$ under the assumption of a free space propagation model is given by
\begin{equation}
    a_{\ell d}[n] = \frac{1}{\lVert \mathbf y_\ell - \mathbf x_d \rVert} ~ \mathrm e^{-j 2 \pi \sfrac{\lVert \mathbf y_\ell - \mathbf x_d \rVert}{\lambda[n]}}.
    \label{eq:idealcoeffs}
\end{equation}

Note that the wavelength $\lambda[n]$ is subcarrier-specific, leading to a subcarrier-dependent phase rotation, which is due to the propagation delay.
The ideal channel coefficients computed according to (\ref{eq:idealcoeffs}) are collected in the matrix
\[
    \mathbf A[n] = \left[ a_{\ell d}[n] \right]_{\ell d} \in \mathbb C^{L \times D}.
\]

\subsection{Estimation of Phase and Gain Error Vectors $\mathbf g[n]$}
\label{sec:subcarrierspecific}
The phase and gain errors of the receiver antennas for one particular subcarrier $n$ are modelled using the vectors $\mathbf g[n] = (g_1[n], \ldots, g_L[n])^\mathrm{T} \in \mathbb C^L$.
This is in contrast to \cite{pierre1991experimental}, which uses a matrix $\mathbf G$, where off-diagonal entries represent mutual coupling between antennas.
Here, we assume that mutual coupling is negligible and therefore only consider gain and phase offsets.
The objective of this step is to estimate $\mathbf g[n]$ from $\mathbf A[n]$ and $\mathbf R[n]$.
To simplify the notation, we will omit the subcarrer index $n$ in the following paragraphs.

Calibration is performed by comparing the ideal channel coefficient vector $\mathbf A_{:d}$ for each time instance $d$ to the actually received vector $\mathbf R_{:d}$.
Assuming strong \ac{LoS} channels between \ac{MOBTX} and \ac{AARX}, the received vectors after calibration should approximately correspond to the ideal vectors except for a random phase rotation due to the transmitter's unknown starting phase, modelled as $s_d = \mathrm e^{j \varphi_d} \in \mathbb C$.
The normalization $|s_d| = 1$ is justified by the approximately constant transmit power throughout all time instances.
With enough measurements under \ac{LoS} conditions, it should be possible to perform accurate phase offset estimation, since, in a sufficiently rich scattering environment, datapoints without \ac{LoS} paths effectively exhibit random phase and gain values.
With $\mathbf s = (s_1, \ldots, s_D)^\mathrm{T}$, this intuition is mathematically formalized as
\begin{equation}
    \mathbf R = \diag ( \mathbf g ) ~ \mathbf A ~ \diag ( \mathbf s) + \mathbf Z,
    \label{eq:model}
\end{equation}
where $\mathbf Z \in \mathbb C^{L \times D}$ is a matrix of additive residual errors with unknown distribution.
In the following, two \emph{least squares} approaches are proposed to find $\mathbf g$ based on (\ref{eq:model}).
The first approach in Sec. \ref{sec:coorddescent} directly transforms (\ref{eq:model}) into an optimization problem that is solved by coordinate descent.
The second approach in Sec. \ref{sec:eigenvec}, on the other hand, does not estimate $\mathbf g$ directly, but first estimates $\mathbf g \mathbf g^\mathrm{H}$, at which point estimating $\mathbf g$ can be expressed as an eigenvalue problem.

\subsubsection{``Coordinate Descent''-Based Least Squares Optimization}\footnote{The authors would like to thank Marvin Geiselhart for his valuable contributions to our investigation of this optimization method.}
\label{sec:coorddescent}
By minimizing the sum of squared absolute residuals $\lVert \mathbf Z \rVert_\mathrm{F}^2$ w.r.t. all unknowns, we obtain an optimization problem:
\begin{align}
    \min_{\substack{\mathbf g, \mathbf s \\ |s_d| = 1}} \left\lVert \mathbf Z \right\rVert_\mathrm{F}^2
    &= \min_{\substack{\mathbf g, \mathbf s \\ |s_d| = 1}} \lVert \diag(\mathbf g) ~ \mathbf A ~ \diag(\mathbf s) - \mathbf R \rVert_\mathrm{F}^2 \nonumber \\
    &= \min_{\substack{\mathbf g, \mathbf s \\ |s_d| = 1}} \sum_{d=1}^D \sum_{\ell=1}^L |g_\ell a_{\ell d} s_d - r_{\ell d}|^2
    \label{eq:optimizationproblem}
\end{align}
In the general case, the quartic optimization problem in (\ref{eq:optimizationproblem}) is not convex.
However, we found that for measured datasets, our algorithm always converges to the same result despite random initialization (except for global phase rotations).

We propose to solve (\ref{eq:optimizationproblem}) using a \emph{coordinate descent} approach.
That is, the objective function is minimized w.r.t. just one coordinate, either $g_\ell$ ($\ell = 1, \ldots, L$) or $s_d$ ($d = 1, \ldots, D$), at a time while all other coordinates are fixed.
Here, coordinate descent leads to closed-form equations for coordinate blocks $\mathbf g$ and $\mathbf s$, within which all coordinates can be updated simultaneously, making it particularly computationally efficient.


\begin{enumerate}[label=(\roman*)]
    \item \textbf{Optimize w.r.t. $g_i$ with all other coordinates fixed}

    Since $|g_\ell a_{\ell d} s_d - r_{\ell d}|^2$ is constant w.r.t. $g_i$ for $\ell \neq i$:
\begin{align*}
    \hat g_i &= 
    \argmin_{g_i \in \mathbb C} \sum_{d=1}^D |g_i a_{i d} s_d - r_{i d}|^2 \\
    &= \argmin_{g_i \in \mathbb C} ~ \lVert g_i ~ (\mathbf A_{i:} \odot \mathbf s) - \mathbf R_{i:} \rVert^2
\end{align*}

This is a linear least squares problem with solution:
\[
    \hat g_i = \frac{(\mathbf A_{i:} \odot \mathbf s)^\mathrm{H}}{\lVert \mathbf A_{i:} \odot \mathbf s \rVert^2} ~ \mathbf R_{i:}
\]

All $g_\ell$ can be optimized simultaneously as a block, since updating $g_i$ does not impact the update for $g_\ell$ with $\ell \neq i$.

    \item \textbf{Optimize w.r.t $s_i$ with all other coordinates fixed}

    Since $|g_\ell a_{\ell d} s_d - r_{\ell d}|^2$ is constant w.r.t. $s_i$ for $d \neq i$:
    \begin{align*}
        \hat s_i &= \argmin_{s_i, |s_i| = 1} \textstyle \sum_{\ell = 1}^L |g_\ell a_{\ell i} s_i - r_{\ell i}|^2 \\
        &= \argmin_{s_i, |s_i| = 1} \left\lVert (\mathbf g \odot \mathbf A_{:i}) ~ s_i - \mathbf R_{:i} \right\rVert^2 \\
        &= \argmin_{s_i} \left\{ - 2 \mathrm{Re} \left\{ \mathbf R_{:i}^\mathrm{H} ~ (\mathbf g \odot \mathbf A_{:i}) ~ \mathrm e^{j \varphi_i} \right\} \right\} 
    \end{align*}
    Omitting the constant factor yields
    \begin{align*}
        \hat s_i 
        &= \argmax_{s_i} \left\{ \cos \left( \arg(\mathbf R_{:i}^\mathrm{H} ~ (\mathbf g \odot \mathbf A_{:i})) + \varphi_i \right) \right\},
    \end{align*}
    which is maximized for $\varphi_i = -\arg(\mathbf R_{:i}^\mathrm{H} (\mathbf g \odot \mathbf A_{:i}))$ and
    \begin{equation}
        \hat s_i = \mathrm e^{-j \arg(\mathbf R_{:i}^\mathrm{H} (\mathbf g \odot \mathbf A_{:i}))}.
        \label{eq:svector}
    \end{equation}

    All $s_d$ can be optimized simultaneously as a block, since updating $s_i$ does not impact the update for $s_\ell$ with $\ell \neq i$.
\end{enumerate}
These two coordinate block optimization steps are summarized in the following iterative algorithm, which estimates $\hat g$ based on $\mathbf R$ and $\mathbf A$ in $M$ iteration steps:

\begin{algorithm}
\caption{Iterative estimation of $\mathbf g$}
\label{alg:iterative}

\KwData{Received / ideal channel coefficients $\mathbf R$ and $\mathbf A$}
\KwResult{Phase and gain offset estimate $\mathbf {\hat g} = \mathbf {\hat g}^{(M)}$}

Randomly initialize $\mathbf { \hat g }^{(0)}$ drawing from $\mathcal N(0, \mathbf I)$\;

\For{$m \gets 1$ \KwTo $M$}{
    \For{$i \gets 1$ \KwTo $D$}{
        $\hat s_i^{(m)} \gets \mathrm e^{-j \arg(\mathbf R_{:i}^\mathrm{H} ~ (\mathbf g^{(m-1)} \odot \mathbf A_{:i}))}$\;
    }
    \For{$i \gets 1$ \KwTo $L$}{
        $\hat g_i^{(m)} \gets \frac{(\mathbf A_{i:} \odot \mathbf {\hat s}^{(m)})^\mathrm{H}}{\lVert \mathbf A_{i:} \odot \mathbf s^{(m)} \rVert^2} ~ \mathbf R_{i:}$\;
    }
}
\end{algorithm}
The choice of a suitable number of iterations $M$ or some other break condition depends on the desired precision.

\subsubsection{Autocorrelation and Principal Eigenvector-Based Estimation}
\label{sec:eigenvec}

The coordinate descent approach in Sec. \ref{sec:coorddescent} needs several iterations to approximate the optimum when estimating $\mathbf g$ directly.
An alternative, less computationally complex approach estimates the autocorrelation matrix $\mathbf g \mathbf g^\mathrm{H}$ and, from it, obtains an estimate for $\mathbf g$.
Solving (\ref{eq:model}) for $\mathbf g \mathbf g^\mathrm{H}$ yields
\begin{align*}
    \mathbf g \mathbf g^\mathrm{H} 
    &= (\mathbf R_{:d} \odot \mathbf A_{:d}^{-1}) (\mathbf R_{:d} \odot \mathbf A_{:d}^{-1})^\mathrm{H} + \mathbf { \tilde Z }_d \quad (d = 1, \ldots, D)
\end{align*}
with residual noise matrices $\mathbf { \tilde Z }_d$ with unknown distribution.
An estimate $\mathbf {\hat C}$ for $\mathbf g \mathbf g^\mathrm{H}$ can be obtained simply by computing the sample mean over all time instances $d$:
\[
    \mathbf {\hat C} = \frac{1}{D} \sum_{d = 1}^D (\mathbf R_{:d} \odot \mathbf A_{:d}^{-1}) (\mathbf R_{:d} \odot \mathbf A_{:d}^{-1})^\mathrm{H}
\]
\vspace{0.2cm}

Note that $\mathbf {\hat C}$ is Hermitian, i.e., $\mathbf {\hat C}^\mathrm{H} = \mathbf {\hat C}$ and, being a sum of positive semidefinite matrices, also positive semidefinite.
To obtain an estimate $\hat {\mathbf g}$ for $\mathbf g$, we apply the least squares method:
\begin{align*}
    \hat {\mathbf g} &= \argmin_{\bm \vartheta \in \mathbb C^L} \lVert \mathbf {\hat C} - \bm \vartheta \bm \vartheta^\mathrm{H} \rVert_\mathrm{F}^2 \\
    &= \argmax_{\bm \vartheta \in \mathbb C^L} \bm \vartheta^\mathrm{H} \mathbf {\hat C} \bm \vartheta + (\bm \vartheta^\mathrm{H} \mathbf {\hat C}^\mathrm{H} \bm \vartheta) - \lVert \bm \vartheta \bm \vartheta^\mathrm{H} \rVert_\mathrm{F}^2
\end{align*}
Since $\mathbf {\hat C}$ is Hermitian, $\bm \vartheta^\mathrm{H} \mathbf {\hat C} \bm \vartheta + (\bm \vartheta^\mathrm{H} \mathbf {\hat C}^\mathrm{H} \bm \vartheta) = 2 \bm \vartheta^\mathrm{H} \mathbf {\hat C} \bm \vartheta \in \mathbb R$, and knowing $\lVert \bm \vartheta \bm \vartheta^\mathrm{H} \rVert_\mathrm{F}^2 = \lVert \bm \vartheta \rVert^4$, we get:
\begin{equation}
    \hat {\mathbf g} = \argmax_{\bm \vartheta \in \mathbb C^L} 2 \bm \vartheta^\mathrm{H} \mathbf {\hat C} \bm \vartheta - \lVert \bm \vartheta \rVert^4
    \label{eq:eigenvecopt}
\end{equation}
By setting the derivative w.r.t. $\bm \vartheta$ to $\mathbf 0$, we obtain
\begin{equation}
    \mathbf { \hat C } \bm \vartheta = \lVert \bm \vartheta \rVert^2 \bm \vartheta
    \label{eq:eigenvec},
\end{equation}
which is a necessary condition for an optimum.
Clearly, (\ref{eq:eigenvec}) is fulfilled by all vectors $\bm \vartheta$ that are appropriately scaled eigenvectors of $\mathbf {\hat C}$.
Namely, they need to be scaled according to their corresponding eigenvalue $\lambda \geq 0$ (since $\mathbf {\hat C}$ is positive semidefinite), such that $\lVert \bm \vartheta \rVert = \sqrt{\lambda}$.
From (\ref{eq:eigenvecopt}) we see that the objective function is maximized if the eigenvector $\bm \vartheta$ with the largest corresponding eigenvalue $\lambda$ is chosen, i.e., the principal eigenvector.
Hence, the estimate $\hat { \mathbf g }$ is chosen out of the set $\left\{ (\bm \vartheta_\ell, \lambda_\ell) \right\}$ of eigenvectors with corresponding eigenvalues of $\mathbf {\hat C}$ and scaled accordingly:
\[
    \ell_\mathrm{princ} = \argmax_\ell \{ \lambda_\ell \} ~~ \rightarrow ~~ \mathbf { \hat g } = \sqrt{\lambda_{\ell_\mathrm{princ}}} ~ \frac{\bm \vartheta_{\ell_\mathrm{princ}}}{\lVert \bm \vartheta_{\ell_\mathrm{princ}} \rVert}
\]
This motivates the following algorithm:

\begin{algorithm}
\caption{Eigenvector-based estimation of $\mathbf g$}
\label{alg:eigenvec}

\KwData{Received / ideal channel coefficients $\mathbf R$ and $\mathbf A$}
\KwResult{Phase and gain offset estimate $\mathbf {\hat g}$}

$\mathbf {\hat C} \gets \frac{1}{d} \sum_{d = 1}^D (\mathbf R_{:d} \odot \mathbf A_{:d}^{-1}) (\mathbf R_{:d} \odot \mathbf A_{:d}^{-1})^\mathrm{H}$\;
Find set of eigenvectors / eigenvalues $\{ (\vartheta_\ell, \lambda_\ell) \}$ of $\mathbf {\hat C}$\;
$\ell_\mathrm{princ} \gets \argmax_\ell \left\{ \lambda_\ell \right\}$\;
$\mathbf { \hat g } \gets \sqrt{\lambda_{\ell_\mathrm{princ}}} ~ \frac{\bm \vartheta_{\ell_\mathrm{princ}}}{\lVert \bm \vartheta_{\ell_\mathrm{princ}} \rVert}$\;

\end{algorithm}

\subsection{Estimation of Time Offsets $t_{\mathrm{off}, \ell}$ and Phase Offsets $\varphi_{\mathrm{off}, \ell}$}
\label{sec:estimatephasetime}
Knowing $\mathbf g[n] \in \mathbb C^L$ for all subcarriers $n$, suitable values for $\varphi_{\mathrm{off, \ell}}$ and $t_{\mathrm{off}, \ell}$ can be estimated based on (\ref{eq:sto}).
We use Kay's single frequency estimator \cite{kay1989fast}, which is robust against noisy phase values including phase jumps by $2\pi$, to obtain an estimate $\hat t_{\mathrm{off}, \ell}$ and then compute an estimate for $\varphi_{\mathrm{off}, \ell}$ using
\[
    \hat \varphi_{\mathrm{off, \ell}} = \arg\left\{\frac{1}{N_\mathrm{sub}} \sum_{n = 1}^{N_\mathrm{sub}} g_\ell[n] ~ \mathrm{e}^{-j 2 \pi n \hat t_{\mathrm{off}, \ell} \Delta f_\mathrm{sub}}\right\}.
\]
\vspace{0.2cm}

\section{Experiments on Measured Datasets}
\label{sec:experiments}
The following analyses will be carried out on two datasets, both of which were captured at a carrier frequency of $1.272\,\mathrm{GHz}$ and with a bandwidth of $50\,\mathrm{MHz}$:
\begin{itemize}
    \item \emph{dichasus-015x}: Indoor office environment datset with $L = 32$ antennas in a single $4 \times 8$ antenna array \cite{dataset-dichasus-015x}.
    \item \emph{dichasus-cf0x}: Indoor factory environment dataset with $L = 32$ antennas arranged in four $2 \times 4$ antenna arrays \cite{dataset-dichasus-cf0x}.
\end{itemize}
\begin{figure}
    \centering
    \includegraphics[width=0.8\columnwidth]{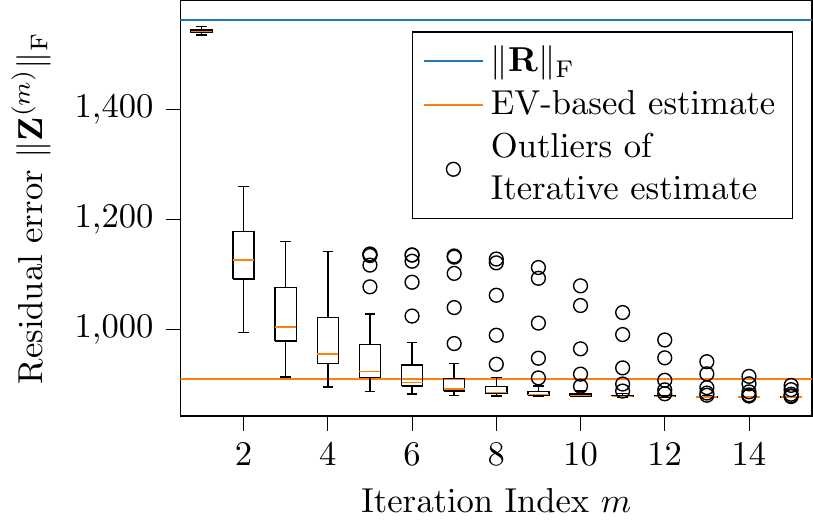}
    \caption{Residual error as a function of the iteration index $m$. The box of the boxplot goes from first to third quantile and the whiskers extend from the box by $1.5\times$ the inter-quartile range.}
    \label{fig:reserror}
\end{figure}
First, we investigate the behavior of the iterative coordinate descent algorithm presented in Sec. \ref{sec:coorddescent} and compare it to the eigenvector-based solution from Sec. \ref{sec:eigenvec}.
We perform this analysis on one particular subcarrier and on a subset of  \emph{dichasus-cf0x} called \emph{dichasus-cf02}.
The results are nevertheless typical for all datasets and subcarriers.
We run the coordinate descent algorithm 40 times, each time with a different initialization $\mathbf g^{(0)}$.
We compute the residual error $\mathbf Z^{(m)}$ after the $m$\textsuperscript{th} iteration step based on (\ref{eq:model}) and visualize the distribution of its Frobenius norm using a boxplot in Fig. \ref{fig:reserror}.
For comparison, we also draw the residual error achieved by the eigenvector-based estimation method (Sec. \ref{sec:eigenvec}) and the value of $\lVert \mathbf R \rVert_\mathrm{F}$, which can be interpreted as the ``worst-case'' residual error that would occur if no calibration was possible.
Clearly, the residual error almost always converges to a minimum after a few iterations.
The result of the eigenvector-based estimation is only slightly worse than the solution found through coordinate descent.

\begin{figure}
    \centering
    \begin{subfigure}{0.37\textwidth}
        \centering
        \includegraphics[width=\textwidth]{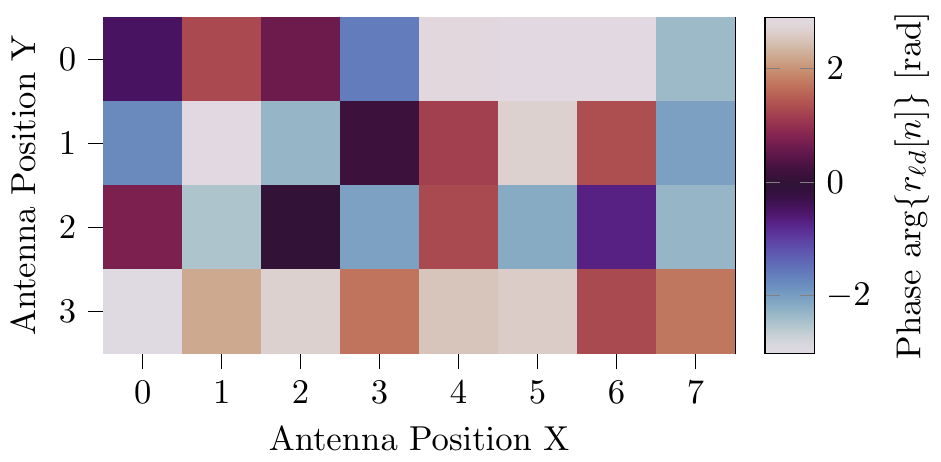}
        \vspace{-0.5cm}
        \caption{Phases $\arg \{ \mathbf R_{:d}[n] \}$ before calibration}
        \label{fig:arrayphasesuncalibrated}
    \end{subfigure}
    \begin{subfigure}{0.37\textwidth}
        \centering
        \vspace{0.0cm}
        \includegraphics[width=\textwidth]{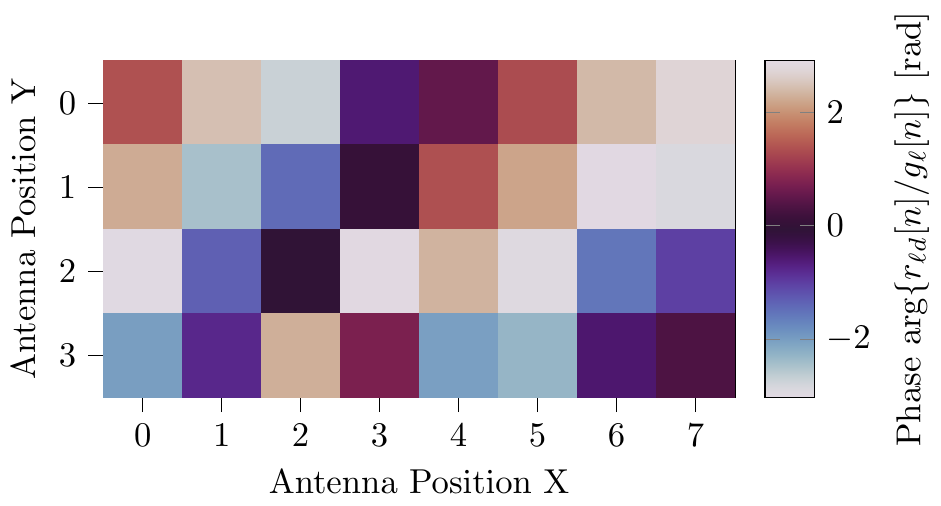}
        \vspace{-0.5cm}
        \caption{Phases $\arg \{ \mathbf R_{:d}[n] \odot \mathbf g[n]^{-1} \}$ after calibration}
        \label{fig:arrayphasescalibrated}
    \end{subfigure}
    \caption{Frontal view: Phases for all array antennas.}
\end{figure}
Second, we visually demonstrate that our calibration makes the phase measurements usable for array processing.
We pick a datapoint from \emph{dichasus-0152}, which is a subset of \emph{dichasus-015x}, where the transmitter is $3\,\mathrm{m}$ in front of, $1.2\,\mathrm{m}$ below and $1.2\,\mathrm{m}$ to the right of the antenna array's center and visualize the received phases over the physical positions of the antennas in the $4\times8$ array.
Before calibration, no pattern is visible in the observed antenna phases (Fig. \ref{fig:arrayphasesuncalibrated}).
After calibration, the phases decrease from bottom right to top left (Fig. \ref{fig:arrayphasescalibrated}), as one would expect for a phase-synchronous array.

\begin{figure}
    \centering
    \begin{subfigure}{0.7\columnwidth}
        \centering
        \begin{tikzpicture}
            \begin{axis}[
                width=0.75\textwidth,
                height=0.75\textwidth,
                scale only axis,
                xmin=-14,
                xmax=4,
                ymin=-17,
                ymax=0,
                xlabel={$x$ coordinate [m]},
                ylabel={$y$ coordinate [m]}
                ]
                    
                \addplot[thick,blue] graphics[xmin=-12,ymin=-14,xmax=2,ymax=-1] {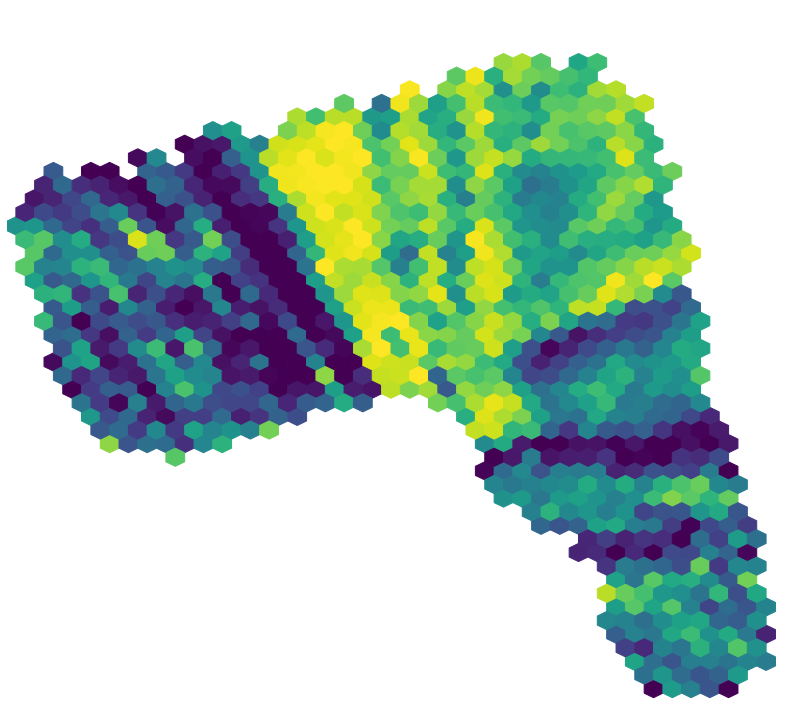};
                \node (arraya) at (axis cs: 2.6747, -13.8973) [draw, fill = black, minimum width = 2, minimum height = 12, inner sep
                = 0pt, rotate = 116.8, anchor = center] {}; 
                \node (arrayb) at (axis cs: -11.250275, -9.689025) [draw, fill = black, minimum width = 2, minimum height = 12, inner sep = 0pt, rotate = 37.2, anchor = center] {}; 
                \node (arrayc) at (axis cs: -1.531375, -15.0595) [draw, fill = black, minimum width = 2, minimum height = 12, inner sep = 0pt, rotate = 77.8, anchor = center] {}; 
                \node (arrayd) at (axis cs: -12.684425, -4.483325) [draw, fill = black, minimum width = 2, minimum height = 12, inner sep = 0pt, rotate = -4.87, anchor = center] {}; 
                
                \node [below = 0.2cm of arraya, xshift = -0.1cm] {A};
                \node [left = 0.1cm of arrayb] {B};
                \node [below = 0.0cm of arrayc, xshift = -0.1cm] {C};
                \node [above = 0.1cm of arrayd, xshift = 0.1cm] {D};
                
                \node at (axis cs: -9, -13) [draw, anchor = center, minimum width = 80, minimum height = 40, inner sep = 0, rotate = 17] {Obstacle};
            \end{axis}
        \end{tikzpicture}
    \end{subfigure}
    \begin{subfigure}{0.15\columnwidth}
        \centering
        \begin{tikzpicture}
            \begin{axis}[
                hide axis,
                scale only axis,
                colormap={reverse viridis}{
                    indices of colormap={
                        \pgfplotscolormaplastindexof{viridis},...,0 of viridis}
                },
                colorbar,
                point meta min=0.2,
                point meta max=1.8,
                colorbar style={
                    height=5cm,
                    width=0.3cm,
                    ytick={0.2,1.0,1.8},
                    ylabel={$\left|s_{\mathrm{C}, d} - s_{\mathrm{B}, d}\right|$}
                }]
                \addplot [draw=none] coordinates {(0,0)};
            \end{axis}
        \end{tikzpicture}
    \end{subfigure}

    \caption{Top view: Locations in \emph{dichasus-cf0x} that contributed to the calibration between arrays C and B}
    \label{fig:contributinglocations}
    \vspace{-0.4cm}
\end{figure}
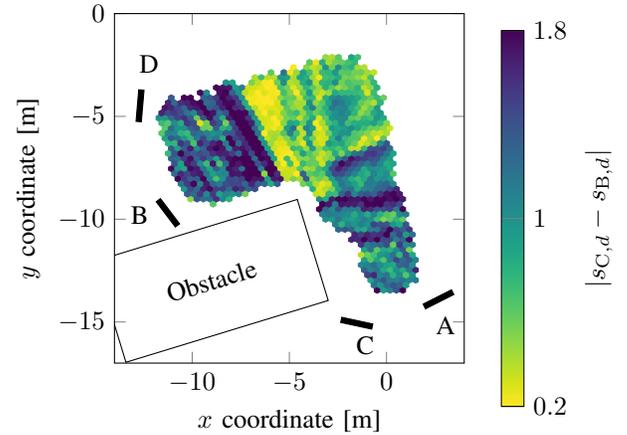
Third, we use \emph{dichasus-cf0x} to visualize that locations with \ac{LoS} paths to spatially distributed antenna arrays are crucial for calibration.
We assign letters A to D to the arrays according to Fig. \ref{fig:contributinglocations} and denote the set of antenna indices belonging to array B and C by $\mathcal B$ and $\mathcal C$, respectively.
We define the metric
\begin{align*}
    \left|s_{\mathrm{C}, d} - s_{\mathrm{B}, d}\right| ~ \text{with} \quad & s_{\mathrm C, d} = \mathrm e^{j \arg\{ \sum_{\ell \in \mathcal C} r_{\ell d}^* g_\ell a_{\ell d} \}} \\
    & s_{\mathrm B, d} = \mathrm e^{j \arg\{ \sum_{\ell \in \mathcal B} r_{\ell d}^* g_\ell a_{\ell d} \}},
\end{align*}
which can be intuitively interpreted in the following way:
$s_{\mathrm{C}, d} \in \mathbb C$ and $s_{\mathrm{B}, d} \in \mathbb C$ are the transmitter starting phases for time instance $d$ estimated according to (\ref{eq:svector}), but \emph{only} based on the antennas that are part of array C or B, respectively.
Then, the absolute value $\left|s_{\mathrm{C}, d} - s_{\mathrm{B}, d}\right|$ is a measure for the agreement between these two starting phase estimates.
If the calibrations $g_\ell$ are valid and a \ac{LoS} channel exists, the starting phases should match.
Visualizing $\left|s_{\mathrm{C}, d} - s_{\mathrm{B}, d}\right|$ for some subcarrier $n$ yields Fig. \ref{fig:contributinglocations}, which demonstrates that locations with \ac{LoS} paths to both array C and B have the largest agreement in starting phase.
This, in turn, implies that these locations contributed to the calibration between arrays B and C.

\section{Performance Comparison of Estimation Algorithms at low SNRs}
\label{sec:performancecomparison}

When deciding which of the the two proposed phase offset estimation algorithms to use, one should, in addition to computational complexity, also consider the accuracy of estimates at operating points with low \acp{SNR}. 
One possible measure for the accuracy of the estimate $\hat {\mathbf g}$ is the squared cosine similarity between $\hat {\mathbf g}$ and the true phase and gain offset vector $\mathbf g$, expressed in logarithmic units:
\[
    P|_\mathrm{dB} = 10 \log_{10} \left( \frac{|\hat {\mathbf g}^\mathrm{H} \, {\mathbf g}|^2}{\lVert \hat {\mathbf g} \rVert^2 \, \lVert \mathbf g \rVert^2} \right) ~ [\mathrm{dB}]
\]
This definition of $P|_\mathrm{dB}$ is invariant to phase rotations equally applied to all components of $\hat {\mathbf g}$, as it should be.
Ideally, $P|_\mathrm{dB} = 0\,\mathrm{dB}$, which is achieved for $\hat {\mathbf g} = \mathbf g$.

\begin{figure}
    \begin{subfigure}{\columnwidth}
        \centering
         \input{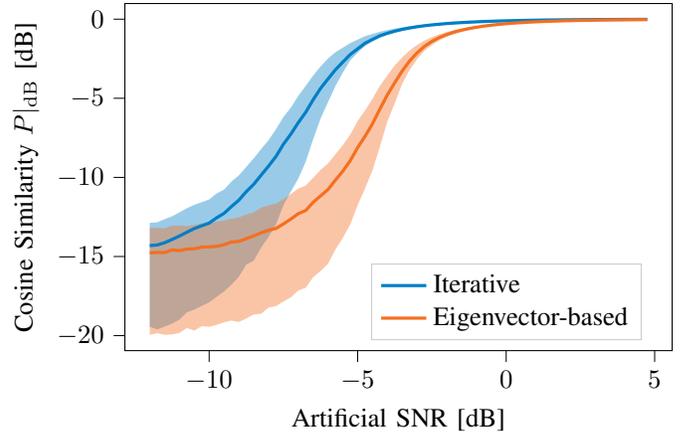}
         \vspace{-0.5cm}
         \caption{Evaluated on dataset \emph{dichasus-cf0x}}
        \label{fig:performanceanalysis-cf0x}
    \end{subfigure}

    \vspace{0.2cm}

    \begin{subfigure}{\columnwidth}
        \centering
        \input{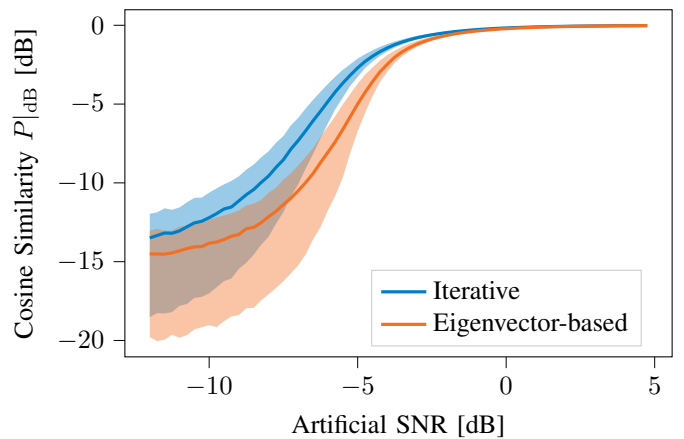}
        \vspace{-0.5cm}
        \caption{Evaluated on dataset \emph{dichasus-015x}}
        \label{fig:performanceanalysis-015x}
    \end{subfigure}
    \caption{Performance of iterative and eigenvector-based phase and gain offset estimation algorithm evaluated at low \acp{SNR}.}
    \label{fig:performanceanalysis}
\end{figure}

To evaluate the performance at low \acp{SNR} on the existing datasets, we only consider a single subcarrier and artificially add white Gaussian noise to the matrix of measured channel coefficients $\mathbf R$.
We compute the \emph{true} vector $\mathbf g$ by estimating phase and gain offsets without artificial noise.
We assume that the estimate without artificial noise is perfectly accurate (without noise, iterative and eigenvector-based estimation generate almost identical estimates).
We then compute $\hat {\mathbf g}$ using both Alg. \ref{alg:iterative} (iterative, with $M = 40$ iterations) and Alg. \ref{alg:eigenvec} (eigenvector-based) with artificial noise powers corresponding to \acp{SNR} between $-12\,\mathrm{dB}$ and $+5\,\mathrm{dB}$ and compute the respective cosine similarities $P|_\mathrm{dB}$.
The result is shown in Fig. \ref{fig:performanceanalysis}: The solid lines correspond to the mean value of $P|_\mathrm{dB}$, computed over 5000 different noise realizations for each \ac{SNR}, and the shaded areas indicate the range from first to third quartile (across random noise realizations).

It can be observed that the iterative algorithm outperforms the eigenvector-based algorithm at low \acp{SNR}.
This result holds for both \emph{dichasus-cf0x} (Fig. \ref{fig:performanceanalysis-cf0x}) and \emph{dichasus-015x} (Fig. \ref{fig:performanceanalysis-015x}), although the performance gap in \emph{dichasus-cf0x} is more pronounced, possibly due to the higher number of datapoints.
While this may not always be the case, it seems that the iterative algorithm is better for most scenarios in terms of estimation accuracy.

\section{Conclusion and Outlook}
\label{sec:conclusion}
A geometry-based calibration procedure was developed and experimentally shown to be applicable to distributed massive \ac{MIMO} channel sounder measurements.
The accuracy of two different calibration algorithms was evaluated for low-\ac{SNR} operating regimes.
If additional information is available, e.g., if receiver gain imbalances are known to be negligible, additional constraints could enhance the calibration's performance.

\bibliographystyle{IEEEtran}
\bibliography{IEEEabrv,references}

\end{document}